\documentclass[twocolumn,twoside,preprintnumbers,amsmath,amssymb,showkeys]{revtex4}
\usepackage{epsfig}
\usepackage{graphicx} 
\usepackage{fancyhdr}
\usepackage{pslatex}

\newcommand{\be}{\begin{equation}}
\newcommand{\ee}{\end{equation}}

\newcommand{\ben} {\begin{eqnarray}}
\newcommand{\een} {\end{eqnarray}}
\newcommand{\nn } {\nonumber}

\def\GeV{\nobreak\,\mbox{GeV}}

\def\pli{p^\prime}

\begin{document}

\title{Coupling constants of $D^* D_s K$ and $D_s ^* D K$ processes}

\author{M. E. Bracco$^{\dag}$}
\author{A. Loz\'ea$^{*}$}
\author{A. Cerqueira Jr.$^\dag$}
\author{M. Chiapparini $^\dag$}
\author{M. Nielsen $^\ddag$}

\affiliation{$^{\dag}$Instituto de F{\'{\i}}sica, Universidade do Estado do Rio de Janeiro, Rua S\~ao Francisco Xavier 524, 20559-900, Rio de
Janeiro, RJ, Brasil.}
\affiliation{$^{*}$Universidade Federal do Rio de Janeiro, C.P. 68528, 21945-970, Rio de Janeiro, RJ, Brasil.}
\affiliation{$^\ddag$ Instituto de F\'{\i}sica, Universidade de S\~{a}o Paulo, 
C.P. 66318, 05389-970 S\~{a}o Paulo, SP, Brazil.}
\received{ on 2006}

\begin{abstract}
We calculate the coupling constants of 
$ D^* D_s K$ and $D_s^* D K$ vertices using the QCD sum rules technique. 
We compare results obtained
in the limit of SU(4) symmetry and found that the symmetry
is broken on the order of 40\%.

\end{abstract}

\maketitle

\thispagestyle{fancy}
\setcounter{page}{0}

The knowledge of coupling constants in hadronic vertices is crucial 
to estimate cross sections when hadronic degrees of freedom
are used. For example absorption of charmonium by kaons will be one of the indicatives 
of the formation of the quark gluon plasma (QGP) in relativistic heavy ions collisions \cite{matsui}.
The kaon is one of the commovers light mesons that can annihilate the charmonium in medium 
given as result D mesons. The quantitative measure of this cross section it is essential to understand
the formation of QGP, but it is not an reality data actually for the experimentalists. 
The processes of absorption of $J/\Psi$ by kaons can be visualized in the Figure 1.
\begin{figure}[b] 
\begin{center}
\epsfysize=4.0cm
\epsfig{file=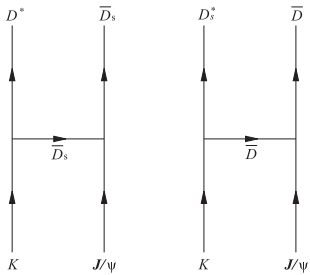}
\caption{Annihilation of $J/\Psi$ by kaons given $Ds$, $D^*$, $D_s^*$ and $D$ mesons production.}
\label{fig1}
\end{center}
\end{figure}
To obtain theoretical values of the cross section, in the case of charmonium 
absorption, one of the more usual approach used is based on effective SU(4) Lagrangians \cite{lin,regina}. 
The effective Lagrangians that described the processes represented in Fig.1 are:
\begin{equation}
{\cal {L}} _{DsD^*K}=i g_{DsD^*K}D^{*\mu}(\bar D_s \partial_{\mu} K
-(\partial_{\mu}\bar D_s) K) +H.c., 
\end{equation}
\begin{equation}
{\cal {L}} _{Ds^*DK}=i g_{Ds^*DK}D_s^{*\mu}(\bar D \partial_{\mu}\bar K
-(\partial_{\mu}\bar D)\bar K) +H.c.. 
\end{equation}
In this formalism it is necessary have the detailed knowledge of the form factors and coupling constants 
in the hadronic vertices to obtain the cross section. 
With the use of the form factor or not, the cross section may change by a factor of two \cite{lin}. 
Also the values of the coupling constants used are calculated by SU(4) exact symmetry and in the case 
the coupling constants are derivate putting the masses of the quarks $u$, $d$ identical to the
$s$ and $c$ quarks. In this way the values of the coupling constants for the two vertex of the right side 
in the processes, in Fig. 1, are identical: 
\begin{equation}
g_{Ds^*DK}=g_{DsD^*K}=\frac{g}{2 \sqrt{2}}.
\end{equation}

In this work we study the $D^* D_s K$ and $D_s ^* D K$ vertices using
the QCD Sum Rules technique \cite{svz}, to obtain the form factors and to infer the 
coupling constants.

We have been working on the problem of compute coupling constants for others
processes and have a consistent method for this 
\cite{ddpi,ddpir,ddrho,matheus1,matheus2,dasilva,psid*d*,australia, dsdk}. Following 
the QCDSR formalism described in our previous works 
\cite{ddpi,ddpir,ddrho,matheus1,matheus2,dasilva,psid*d*,australia, dsdk}, 
we write the three-point correlation function associated 
with the $D^* D_s K$  vertex, which is given by

\ben
\Gamma_{\mu}^{(K)}(p,\pli) & = & \int d^4x \;\; d^4y \;\;
e^{i\pli\cdot x} \; e^{-i(\pli-p)\cdot y} \nn \\
& & \langle 0|T\{j_{\mu}^{D^*}(x) {j^K}^\dagger(y) 
 {j^{D_s}}^\dagger(0)\}|0 \rangle
\label{corkoffnos} 
\een
for $K $ meson off-shell, where the interpolating currents are 
$j_\mu^{D^*}=\bar c\gamma_\mu d$,
$j^K=i\bar s\gamma_5 d$ and
$j^{D_s}=i\bar c\gamma_5 s$, and
\ben
\Gamma_{\mu \nu}^{(D_s)}(p,\pli)& =  & \int d^4x \; 
d^4y \;\; e^{i\pli\cdot x} \, e^{-i(\pli-p)\cdot y}\; \nn \\
& & \langle 0|T\{j_{\mu}^{K}(x)  {j^{D_s}}^\dagger(y) 
 {j_{\nu}^{D^*}}^\dagger(0)\}|0\rangle \label{cordsoffnos} 
\een
for $D_s$ meson off-shell, with the interpolating currents 
$j^{K}_\mu = \bar u \gamma_{\mu}\gamma_{5} s$, 
$j^{D_s} = i \bar c \gamma_{5} s$, 
$j_\mu^{D^*}= \bar u \gamma_{\mu} c$, with
$u$, $d$, $s$ and $c$ being the $up$, $down$, $strange$ and $charm$ 
quark fields respectively. In both cases, each one of these currents has 
the same quantum numbers as the corresponding mesons.

Using the above currents to evaluate the correlation functions 
(\ref{corkoffnos}) and (\ref{cordsoffnos}), the theoretical or QCD side 
is obtained.  The framework to calculate the correlators in the QCD side 
is the Wilson operator product expansion (OPE). The Cutkosky's rule 
allows us to obtain the double discontinuity of the correlation function 
for each one of the Dirac structures appearing in the correlation function.
Then we use spectral representation over the virtualities $p^2$ and ${\pli}^2$,
 holding $Q^2= -q^2$ fixed. The amplitudes receive contributions  from 
all terms in the OPE. The leading contribution comes from 
the perturbative term.

The phenomenological side of the sum rule, which is written in terms of 
the mesonic degrees of freedom, is parametrized in terms  of the form 
factors, meson decay constants and meson masses. We introduce the meson decay constants 
$f_{K}$, $f_{D_s}$ and $f_{D^*}$, which are defined by the following matrix 
elements 
\begin{equation}
\langle 0|j^{K}|{K}\rangle=\frac{m_{K}^2 f_{K}}{m_s+m_q},
\end{equation}
\begin{equation}
\langle 0|j^{D_s}|{D_s}\rangle=\frac{m_{D_s}^2}{m_c+m_s}f_{D_s}
\end{equation}
 and
\begin{equation}
\langle 0|j_{\nu}^{D^*}|{D^*}\rangle=m_{D^*}f_{D^*} \epsilon^*_{\nu},
\end{equation}
where $\epsilon_{\nu}$ is the polarization vector of the $D^*$ meson. 
 The QCD sum rule is obtained by matching both representations, using the universality 
principle. The matching is improved by performing a double Borel 
transform on both sides. The perturbative contribution for both Eqs.~(\ref{corkoffnos})
 and (\ref{cordsoffnos}) is given in details in ref.\cite{dsdk}.
We chosen one structure that appear in both sides and also
 must have a stability that guarantees a good match between the two sides of the sum rule. 
The structures that obey these two points are $\pli_{\mu}$, in the case
$K$ off-shell, and $\pli_{\mu}\pli_{\nu}$ in the case $D_s$ off-shell. 

The Borel transformation \cite{io2} in the variables 
$P^2=-p^2\rightarrow M^2$ and $P'^2=-{\pli}^2\rightarrow M'^2$ allows to get 
the final form of the sum rule, which allow us to obtain the form factors 
$g^{(M)}_{D^* D_s K}(Q^2)$  where $M$ stands for the off-shell meson. 

We use Borel masses satisfying the constraint 
$M^2/M'^2=m_{in}^2/m_{out}^2$, where $m_{in}$ and 
$m_{out}$ are the masses of the incoming and out coming meson respectively.
The values of the parameters used in the calculation of the vertices are 
depicted in Table~\ref{param1} and in Table ~\ref{param2}
\begin{table}[t]
\begin{tabular}{|c|c|c|c|c|c|c|c|c|}\hline
$m_q$&$m_s$&$m_c$&$m_K$&$m_{D_s}$&$ m_{D_s^*}$ &$ m_{D}$&$m_{D^*}$\\ 
\hline\hline
0.0&0.13&1.2&0.498&1.97&2.11&1.87&2.01\\ \hline
\end{tabular}
\caption{Masses of quarks and mesons used in the calculation of the QCD sum rule. 
All quantities are in $\GeV$.}\label{param1}
\end{table}

\begin{table}[t]
\begin{tabular}{|c|c|c|c|c|}\hline
$f_K$\cite{fk}&$f_{D_s}$\cite{fds}&$f_{D^*}$\cite{fd*} &$f_{D}$\cite{fd}&$f_{D_s^*}$\cite{fds*}\\ 
\hline\hline
0.160&0.280&0.240&0.200&0.330\\ \hline
\end{tabular}
\caption{Decay constant used in the calculation of the QCD sum rule. 
All quantities are in $\GeV$.}\label{param2}
\end{table}

The continuum thresholds $s_0$ and $u_0$ are the two parameters 
that are including in the QCDSR, and are important to controlate
the pole contribution. Using $\Delta_s=\Delta_u = 0.5 \GeV $ for the continuum thresholds 
and fixing $Q^2=1\GeV^2$, we found a good stability of the form factor
$g_{D^*D_sK}^{(K)}$, as a function of the Borel mass $M^2$, in the 
interval $3< M^2 < 5 \GeV^2$. In the case of the form factor $g_{D^*D_sK}^{(D_s)}$ 
the interval for stability of the sum rule is $2<M^2<5 \GeV^2 $. 

\begin{figure}[t] 
\centering
\includegraphics[scale=0.80]{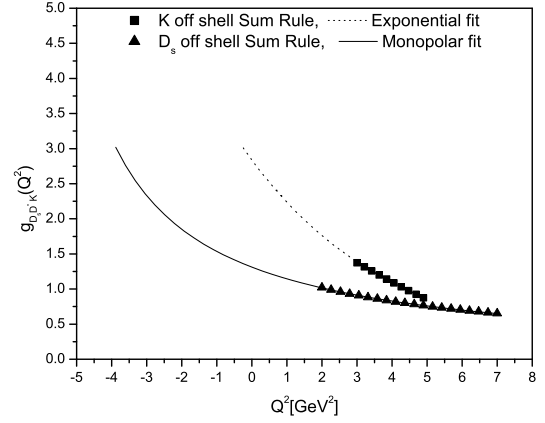}
\caption{$g^{(K)}_{D^*D_sK}$ (squares) and $g^{(D_s)}_{D^*D_s K}$ 
(triangles) form factors as a function of $Q^2$ from the QCDSR 
calculation of this work. The solid (dotted) line corresponds to the 
monopole (exponential) parametrization of the QCDSR results for each case.}
\label{fig2}
\end{figure}

Fixing $\Delta_s=\Delta_u=0.5 \GeV$ and $M^2=3 \GeV ^2$, we evaluate 
the momentum dependence of both form factors. The results are shown in 
Fig.~\ref{fig2}, where the squares corresponds to the 
$g_{D^*D_s K}^{(K)}(Q^2)$ form factor in the  interval where the 
sum rule is valid. The triangles are the result of the sum rule for the 
$g_{D^*D_s K}^{(D_s)}(Q^2)$ form factor. 

In the case of the $K$ meson off-shell, our numerical results can be 
parametrized by an exponential function (dotted line in Fig.~\ref{fig2}):
\begin{equation}
g_{D^* D_s K}^{(K)}(Q^2)= 2.83 \; e^{-\frac{Q^2}{4.19}} \rightarrow g_{D^* D_s K}^{(K)}= 3.01,
\label{expnos}
\end{equation}
where the define coupling constant, $g_{D^* D_s K}^{(K)}$ is chosen as the value of 
the form factor at $Q^2=-m^2_K$. 

In the case when the $D_s$ meson is off-shell, our sum rule results can 
be parametrized by a monopole formula (solid line in Fig.~\ref{fig2}):  
\begin{equation}
g_{D^* D_s K}^{(D_s)}(Q^2)=\frac{9.01}{Q^2+6.86} \rightarrow g_{D^* D_s K}^{(D_s)}= 3.02,
\label{mononos}
\end{equation}
where $g_{D^* D_s K}^{(D_s)}$ is the coupling constant is chosen as the value of 
the form factor at $Q^2=-m^2_{D_s}$.

Comparing the results in Eqs.(\ref{expnos}) and (\ref{mononos}) we see 
that the method used to extrapolate the QCDSR results in both cases, 
$K$ and $ D_s$ off-shell, allows us to extract values for the coupling 
constant which are in very good agreement with each other.

In order to study the dependence of this results with the continuum 
threshold, we vary $\Delta_s=\Delta_u$ in the interval
$0.4 \le \Delta_s=\Delta_u \le 0.6~\GeV$. 
This procedure give us uncertainties in such a way that the final results for the couplings in each case are: 
$$g_{D^* D_s K}^{(K)}= 3.02 \pm 0.15 $$ and $$g_{D^* D_s  K}^{(D_s)}= 
3.03 \pm 0.14. $$

Now we study the $D_s^* D K$ vertex. The treatment is similar to 
the previous case. For details of the calculation see reference 
\cite{dsdk}. 
The correlation functions are 
\ben
\Gamma_{\mu}^{(K)}(p,\pli)&=& \int d^4x \, d^4y \;\;
e^{i\pli\cdot x} \, e^{-i(\pli-p)\cdot y} \nn \\
& & \langle 0|T\{j_{\mu}^{D^*_s}(x) {j^{K}}^\dagger(y) 
 {j^{\bar D}}^\dagger(0)\}|0\rangle  \label{corkoffang} 
\een
for $K$ meson off-shell, where the interpolating currents are 
$j_\mu^{D_s^*}=\bar c\gamma_\mu s$,
$j^K=i\bar u\gamma_5 s$ and
$j^D=i\bar c \gamma_5 u$, 
and 
\ben
\Gamma_{\mu \nu}^{(D)}(p,\pli)&=&\int d^4x \, 
d^4y \;\; e^{i\pli\cdot x} \, e^{-i(\pli-p)\cdot y}\; \nn \\
& & \langle 0|T\{j_\mu^K(x)  {j^D}^\dagger(y)
 {{j_\nu}^{D_s^*}}^\dagger(0)\}|0\rangle 
\label{cordoffang} 
\een
for $D$ meson off-shell, with the interpolating currents 
$j^{K}_\mu = \bar u \gamma_\mu\gamma_5 s$,
$j_{\nu}^{D^*_s}= \bar c \gamma_{\nu} s$, and 
$j^{D} = i \bar u \gamma_{5} c$.
We introduce the decay constants 
$f_D$ and $f_{D_s^*}$, which are defined by the following matrix elements:
\ben
\langle 0|j^{D}|D\rangle&=&\frac{m_D^2}{m_c+m_q}f_D, \label{fd} \\ 
\langle 0|j_{\nu}^{D_s^*}|{D_s^*}\rangle&=&m_{D_s^*}f_{D_s^*} 
\epsilon^*_{\nu},\label{fds*}
\een
where $\epsilon_{\nu}$ is the polarization vector of the $D_s^*$ meson.

\begin{figure}[t] 
\centering
\includegraphics[scale=0.80]{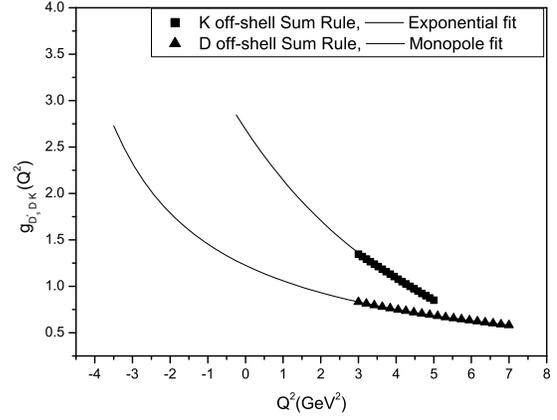}
\caption{$g^{(K)}_{D^*_s D K}$ (squares) and 
$g^{(D)}_{D^*_s D K}$ (triangles) form factors as a function of
$Q^2$ from the QCDSR calculation of this work. The dashed (solid) line 
corresponds to the exponential (monopole) parametrization of the QCDSR 
results for each case.}
\label{fig3}
\end{figure}

In Fig.~\ref{fig3} the squares corresponds to the 
$g_{D^*_s D K}^{(K)}(Q^2)$ form factor in the  interval where the 
sum rule is valid. The triangles are the result of the sum rule for the 
$g_{D^*_s D K}^{(D)}(Q^2)$ form factor.

In the case when the $K$ meson is off-shell, our numerical results can be 
parametrized by an exponential function (dashed curve in Fig.~\ref{fig3})
and the coupling constant is extracted at the value of 
the form factor at $Q^2=-m^2_{K}$:
\begin{equation}
g_{D^*_s D K}^{(K)}(Q^2)= 2.69 \; e^{-\frac{Q^2}{4.39}}\rightarrow g_{D^*_s D K}^{(K)}= 2.87.  
\label{expang}
\end{equation}
In the case when the $D$ meson is off-shell, the sum rule results are
represented by the triangles in 
Fig.~\ref{fig3}, and they can be parametrized by a monopole formula
(solid line in the figure) and the coupling constant is the value of 
the form factor at $Q^2=-m^2_{D}$:  
\begin{equation}
g_{D^*_s D K}^{(D)}(Q^2)=\frac{7.78}{Q^2+6.34} \rightarrow g_{D^*_s D K}^{(D)}= 2.72.
\label{monoang}
\end{equation}

Studying the dependence of our results with the continuum threshold, for 
$\Delta_{s,u}$ varying in the interval  $0.4 \le \Delta_{s,u}\le 0.6~\GeV$,
 we obtain the following values, with errors, for the couplings in each case: 
 $$g_{D^*_s D K}^{(K)}= 2.87 \pm 0.19 $$
 and $$g_{D^*_s D K}^{(D)}= 2.72 \pm 0.31.$$

Concluding, we have studied the form factors and coupling constants of 
$D^* D_s K$ and 
$D_s ^* D K$ vertices in a QCD sum rule calculation. For each case we 
have considered two particles off-shell, the lightest and one of the 
heavy ones: the $K$ and $D_s$ mesons for the $D^*D_sK$ vertex, and the 
$K$ and $D$ mesons for the $D_s ^* D K$ vertex. 
In the two situations, the off-shell particles give compatible results 
for the coupling constant in each vertex.
The results are :
\begin{equation}
g_{D^* D_s K}= 3.02 \pm 0.14 
\label{eqgf1}
\end{equation}
and
\begin{equation}
g_{D_s ^* D K}= 2.84 \pm 0.31.
\label{eqgf2}
\end{equation}
We can compare our result with the prediction of the exact SU(4) 
symmetry \cite{regina}, which would 
give the following relation among these numbers \cite{regina}: $g_{D^* D_s K}=
g_{D_s ^* D K} = 5.$
 Eqs.~(\ref{eqgf1}) and ~(\ref{eqgf2}) shows that the coupling constants 
in the vertices $D^* D_s K$ and $D_s ^* D K$ are consistent one with the  
other, but that they are relatively 
far from the value given by the SU(4) symmetry in the cited reference. Therefore, 
we conclude that the  SU(4) symmetry is broken by 
approximately 40\% in the calculation performed here. 
This is expected because the coupling constant obtain by the exact SU(4) symmetry put
 the masses of $u$ $d$ quarks are the same values that the $s$ and $c$ quarks.
In this case there is not experimental value to compare our result but the values
of the coupling constants were obtained by two different way at the same vertex and
give compatible results.

\end{document}